# Efficiency and Cost Optimization of Dual Active Bridge Converter for 350kW DC Fast Chargers

Sadik Cinik
*Department of Energy*
*Aalborg University*
Aalborg, Denmark
sci@energy.aau.dk

Fangzhou Zhao
*Department of Energy*
*Aalborg University*
Aalborg, Denmark
fzha@energy.aau.dk

Giuseppe De Falco
*Infineon Technologies Austria AG*
Villach, Austria
Giuseppe.DeFalco@infineon.com

Xiongfei Wang
*Division of Electric Power and Energy Systems*
*KTH Royal Institute of Technology*
Stockholm, Sweden
xiongfei@kth.se

***Abstract*— This study focuses on optimizing the design parameters of a Dual Active Bridge (DAB) converter for use in 350 kW DC fast chargers, emphasizing the balance between efficiency and cost. Addressing the observed gaps in existing high-power application research, it introduces an optimization framework to evaluate critical design parameters—number of converter modules, switching frequency, and transformer turns ratio—within a broad operational voltage range. The analysis identifies an optimal configuration that achieves over 95% efficiency at rated power across a wide output voltage range, comprising seven 50 kW DAB converters with a switching frequency of 30 kHz, and a transformer turns ratio of 0.9.**

***Keywords*— *DC Fast Charging, Dual Active Bridge Converter, Wide Output Voltage, Efficiency and Cost Optimization***

## I. INTRODUCTION

Over the last decade, there has been a significant increase in electric vehicle (EV) adoption, driven by substantial incentives and investments, despite challenges such as limited range capacities and long charging times [1]. DC fast charging technology, particularly through the deployment of 350 kW ultra-high-power units, is viewed as a key solution, aiming to reduce charging times to below 10 minutes.

To achieve this goal, the industry is shifting from 400V to 800V systems in passenger vehicles to enable high-power charging without increasing current drawn by batteries [2]. This evolution requires DC fast chargers that support a wide output voltage range, from 150V to 1000 V. However, the demands for high power and broad output voltage range, coupled with pre-existing necessities such as galvanic isolation, introduce complex design challenges. These challenges, directly tied to converter technology critical for charging station functionality, include efficiency degradation over wide voltage range operations and increased costs.

Today, DC fast chargers utilize two power conversion stages: an AC/DC unit for AC to DC conversion and a DC/DC unit to regulate this DC for EV battery voltage needs. These chargers integrate isolation features, through a large line-frequency transformer prior to the AC/DC stage or a smaller, more efficient one within the DC/DC converter stage. The latter is favored for its reduced size and weight, providing necessary isolation [3]. Additionally, the DC/DC converter must sustain high efficiency across a broad output voltage range, a critical aspect for optimizing the charging process.

Existing literature, therefore, predominantly focuses on enhancing the efficiency of isolated DC/DC converters across a wide output voltage ranges, with numerous studies demonstrating efficiency improvements through advanced modulation techniques [4] or hardware modifications [5,6]. However, the literature overlook two critical aspects. First, the balance between efficiency and cost is often neglected, which is crucial for ensuring the economic viability of charging systems. Second, while converters are designed and optimized for lower power levels, the trade-offs for high-power scenarios like 350 kW DC fast chargers remain under-investigated. For example, the optimal configuration and number of converter modules in building a 350 kW system require further research.

To address the identified need, this paper aims to extend the isolated DC/DC converter optimization analysis specifically for 350 kW DC fast chargers. It employs a detailed framework to evaluate how various design parameters—such as converter module count, switching frequency, and transformer turns ratio—affect efficiency and cost across different output voltages. For this optimization, the Dual Active Bridge converter is chosen due to its widespread utilization in the industry, simplicity, inherent buck-boost operation capability, and bidirectionality. Section II details the optimization methods and converter modelling, while Section III discusses how these parameters impact efficiency-cost trade-offs in varied voltage operations.

## II. METHODOLOGY

This section presents a systematic optimization procedure, depicted in Fig. 1, aiming to identify the optimal design parameters for the DAB converter. These parameters include the switching frequency, the transformer turns ratio, and the number of converter modules necessary to achieve a total output power of 350 kW. The brute force approach methodically tests all possible combinations of parameters within the predefined ranges to ensure the identification of the configuration that best balances efficiency and cost.

### A. *Overall System Requirements*

The optimization flow begins with the overall system requirements, which cover CHARIN's DC fast charger specifications [7]. It specifies a maximum power ($P_{max}$) of 350 kW and a maximum current ($I_{max}$) of 500 A, with an output voltage range ($V_{out}$) of 150 V to 1000 V.

### B. *Optimization Design Space*

The study considers three critical parameters that significantly impact the cost and efficiency of converters for DC fast charging applications. These parameters are varied within predefined ranges to determine the optimal configuration that balances efficiency with cost-effectiveness:

**Number of Converter Modules (N):** Explored within a range from 5 to 35, this parameter determines how many individual converters are required to be stacked together to reach the target power output of 350 kW.

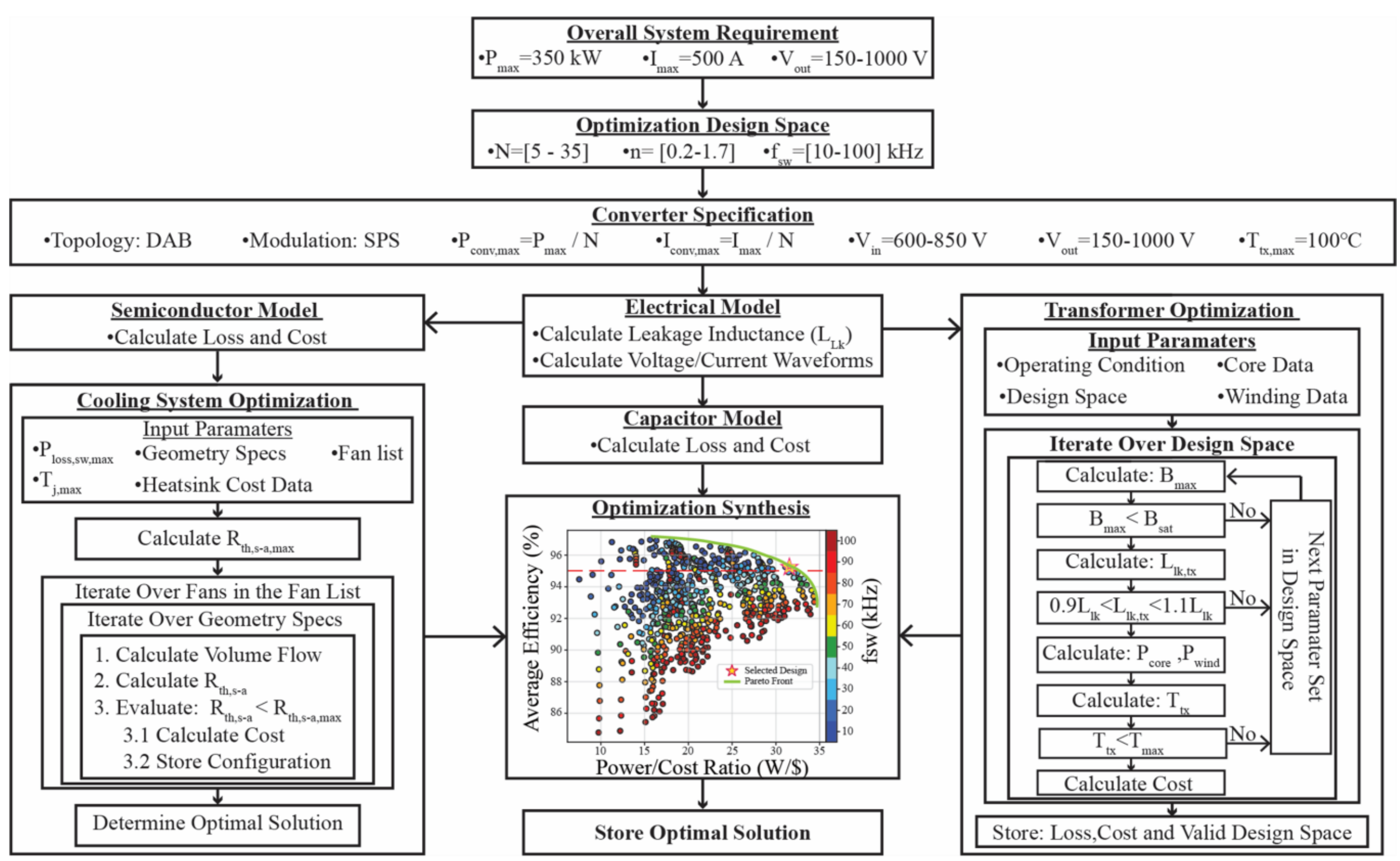

Fig. 1. Flowchart of the optimization process for identifying the optimal design parameters of the DAB converter, including switching frequency, transformer turns ratio, and number of converter modules, to achieve a total power output of 350 kW.

**Switching Frequency ($f_{sw}$):** Analyzed between 10 kHz and 100 kHz, $f_{sw}$ is crucial for the efficiency and cost of the converter components, directly affecting its performance, size, and thermal management.

**Transformer Turns Ratio (n):** Examined from 0.2 to 1.7, n is crucial for voltage transformation, impacting the transformer's cost, efficiency, and design complexity.

## *C. Converter Specification*

This section details the specific parameters and operational principles that define the converter's design within the context of this optimization study.

**Power and Current Ratings:** The power ($P_{conv,max}$) and current ($I_{conv,max}$) ratings for the converter are determined by dividing $P_{max}$ and $I_{max}$ by N.

**Input and Output Voltage Ratings:** The study assumes that input voltage ($V_{in}$) can be regulated between 600 V and 850 V to enhance the converter efficiency under wide output voltage range operation. The output voltage ($V_{out}$) maintained as specified in the system requirements (150 V - 1000 V).

**Temperature Constraints:** The optimization stipulates that the maximum temperature for the transformer ($T_{tx,max}$) should not exceed 100 ℃, and semiconductor junction temperature ($T_{j,max}$) should not exceed the maximum junction temperature specified in the semiconductor datasheet, considering an ambient temperature ($T_{amb}$) of 25 ℃.

**Topology and Modulation:** The DAB converter is chosen for its notable features [4], illustrated in Fig. 2.a, showcasing its dual-active bridge architecture linked by a high-frequency transformer. The bridges are modelled as AC voltage sources in rectangular waveforms, as depicted in Fig. 2.b. The inductor $L_{lk}$, essential for the power transfer, is the transformer's leakage inductance. Its instantaneous current shown in Fig. 2.c is described in (1).

$$\mathrm{i_L(t_i) = i_L(t_{i-1}) + \frac{1}{L_{lk}} \int_{t_{i-1}}^{t_i} \left( v_{ac,1}(t) - \frac{v_{ac,2}(t)}{n} \right) dt} \qquad (1)$$

where $i_L(t_i)$ represents the inductor current at any given time, $L_{lk}$ is the leakage inductance, $v_{ac,1}(t)$ and $v_{ac,2}(t)$ are the transformer's primary and secondary voltages, respectively, and n is the turns ratio. The inductor current $i_L$, within the DAB converter can be modulated by altering the phase shift ratio (D) between the dual full bridges, and the individual phase shifts (D1, D2) within each bridge's legs, offering direct power flow control. While various modulation strategies could refine performance through these phase shifts, this research uses the Single-Phase Shift (SPS) approach for its ease and straightforward implementation.

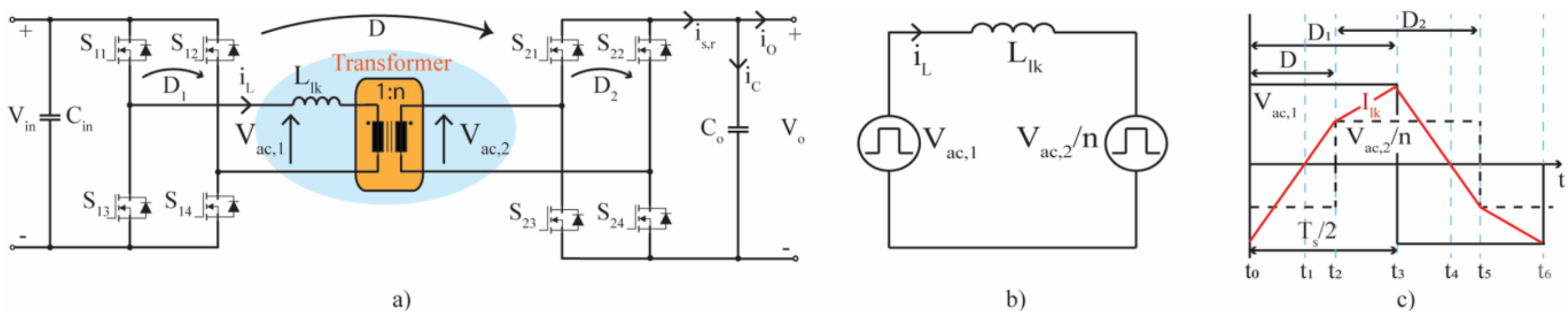

Fig. 2. (a) DAB converter circuit configuration. (b) Equivalent circuit model. (c) Time domain representation of instantaneous inductor current.

### D. Electrical Model

The initial phase in developing the electrical model is to compute the necessary leakage inductance, a key factor impacting subsequent current waveform calculations essential for optimization.

**Leakage Inductance Determination:** The leakage inductance $L_{lk}$ is a key parameter for the DAB converter, directly impacting power delivery capability. To ensure the required rated power can be reached for each output voltage operation, the leakage inductance is computed as:

$$L_{lk} = \frac{V_{out} V_{in} D(1-D)}{2 n f_{sw} P_{conv,max}} \tag{2}$$

**Voltage and Current Waveforms Determination:** The primary and secondary voltage waveforms are defined as square waves, with their amplitudes adjusted based on the input and output voltages. Given the 150 - 1000 V range needed for fast charging, the converter operates in both boost and buck modes. This functionality results in different current waveform patterns in the converter. Utilizing these current waveforms, formed by using (1), (2), enable precise current calculations for any active switch or diode. Such detailed analysis is critical for the subsequent converter component models, directly influencing the overall optimization flow.

### E. Global Optimization Strategy

In the optimization strategy, costs and average operational losses for key converter components—semiconductors, cooling systems, transformers, and capacitors—are evaluated under various design scenarios determined by configurations of $f_{sw}$, N, and n. This comprehensive assessment is conducted using the brute force optimization algorithm, ensuring every possible configuration is considered. The strategy comprises two core analyses for each variable set:

1. Cost Calculations: The costs for each component are determined for every design variable combination to quantify the financial impact of each configuration.
2. Operational Loss Calculations: Component losses are calculated across different voltage (150 V to 1000 V) and power levels (10% and 100% of rated power). By analysing losses at each voltage and power level intersection, a comprehensive profile of the converter's operational behaviour is established. These losses are averaged to establish a benchmark loss metric for each design variable set.

These collective cost and loss evaluations for each design variable set ($f_{sw}$, N, n) are then integrated into an optimization synthesis block. This block consolidates the data, supporting strategic decisions to optimize the converter's cost and operational efficiency, aiming for an overarching optimized solution.

### F. Converter Component Models

This section delves into detailed analyses of the key converter components: semiconductors, cooling systems, transformers, and capacitors.

#### 1) Semiconductor Model

Semiconductor devices are pivotal in power converter designs, influencing both efficiency and cost. This study utilizes 1200 V SiC MOSFETs from Infineon, noted for their reliability and efficiency, detailed in Table I. These devices, selected across various current ratings, accommodate the DAB converter's 10 kW to 70 kW output range, aligned with the N values in design variables. Component pricing, essential for cost analysis, was obtained from Mouser Electronics, reflecting the cost for 100 units as of March 1, 2024, facilitating our cost optimization evaluation.

TABLE I. INFINEON 1200 V SIC MOSFET LIST

| **Part Number** | $R_{ds(on)}$ | $I_d$@100 ℃ | $Cost_{100}$ |
|---|---|---|---|
| IMZA120R007M1H | 7 mΩ | 168 A | $62.39 |
| IMZA120R014M1H | 14 mΩ | 89 A | $32.34 |
| AIMZH120R020M1T | 20 mΩ | 71 A | $29.15 |
| AIMZH120R030M1T | 30 mΩ | 49 A | $20.62 |

Semiconductor losses are mainly categorized into conduction and switching losses. The upcoming subsection outlines the method to calculate these losses, leveraging the semiconductor device specifications from manufacturer datasheets and the converter's operational conditions. These conditions cover switching frequency, ambient temperature, and previously determined current/voltage waveforms.

*a) Conduction Losses:*

Semiconductor characteristics $i_{sw} = f(v_{sat})$ and anti-parallel diode characteristics $i_d = f(v_d)$ at various temperatures are derived from datasheets and organized into lookup tables. These tables, combined with respective current profiles, allow accurate conduction loss calculations for both semiconductors and diodes by interpolating across diverse operating conditions. The total conduction loss is then determined using (3) as outlined in [14].

$$P_{cond} = \frac{1}{T_{sw}} \left( \int_0^{T_{sw}} i_{sw}(t) * v_{sw}\big(i_{sw}(t), T_j\big) dt + \int_0^{T_{sw}} i_d(t) * v_d\big(i_d(t), T_j\big) dt \right) \tag{3}$$

In this formulation, $T_{sw}$ represents the switching period. The $i_{sw}(t)$ and $i_d(t)$ instantaneous currents through the semiconductor switch and diode, while the functions $v_{sw}\big(i_{sw}(t), T_j\big)$ and $v_d\big(i_d(t), T_j\big)$ specify how the saturation voltage of the semiconductor and the forward voltage of the diode vary with their respective currents and junction temperatures ($T_j$).

*b) Switching Losses:*

Switching losses, which occur during the transient states of turn-on and turn-off events, are inherently more complex than steady-state conduction losses. Standardized turn-on ($E_{on}$) and turn-off ($E_{off}$) energy values are provided by manufacturers for typical conditions, which may not reflect the actual operating environment. To bridge this gap, scaling factors for gate resistance ($K_{RG}$), junction temperature ($K_{tj}$), and switching voltage ($K_{V_{sw}}$) are employed to modify these values, thus allowing for an accurate determination of switching losses [14]:

$$P_{sw} = \frac{1}{T_{sw}}\left(E_{on}(I_{sw}) * K_{RG_{on}} * K_{T_{J_{on}}} * K_{V_{sw_{on}}}\right) + \frac{1}{T_{sw}}\left(E_{off}(I_{sw}) * K_{RG_{off}} * K_{T_{J_{off}}} * K_{V_{sw_{off}}}\right) \quad (4)$$

with

$$K_{RG(on/off)} : E_{on/off}(R_G^{user}) / E_{on/off}(R_G^{DS})$$

$$K_{Tj(on/off)} : E_{on/off}(T_j^{user}) / E_{on/off}(T_j^{DS})$$

$$K_{V_{sw(on/off)}} : V_{ce(on/off)}^{user} / V_{ce(on/off)}^{DS}$$

Here, the subscript "DS" denotes datasheet values, while "user" denotes the actual operating conditions. If the diode's reverse recovery energy $E_{rr}$ is not included in $E_{on}$ or $E_{off}$, its reverse recovery loss $P_{rr}$ needs to be calculated using a formula similar to (4). This loss should then be added to $P_{sw}$.

Subsequently, the total loss per switch, $P_{loss,sw,max}$, calculated as the sum of $P_{cond}$ and $P_{sw}$, for each switch listed in Table I, is integrated into the cooling system model to optimize thermal management.

*2) Cooling System Model*

Effective cooling management is vital for semiconductor performance, keeping its junction temperature below the datasheet-specified $T_{j,max}$. While cooling systems are often overlooked in efficiency optimizations due to minimal losses, their significant cost impact necessitates integrating a detailed cooling system model into the optimization process. Inspired by [8] that optimizes the weight of forced convection cooling systems for airborne wind turbines, this study adopts a similar approach to assess and reduce the cooling system's cost, featuring a fan, an adapter, and a straight-fin aluminum heat sink as shown in Fig. 3.b.

Distinct cooling system designs for the primary and secondary side switches of the DAB converter are required due to the potential variance in thermal behaviour and power loss between two bridges. The cooling system optimization outlined in the overall optimization flow chart (Fig. 1) is methodically structured with the following steps.

*a) Definition of Input Parameters*

In addition to the input parameters like $T_{j,max}$, $T_{amb}$, $P_{loss,sw,max}$ already provided previously, the process begins with the definition of critical parameters that include:

**Heatsink Geometry:** Key dimensions defining heatsink design optimization, as illustrated in Fig. 3.a, are detailed with their respective ranges in Table II. These dimensions are based on standard design practices and guidelines detailed in [8].

TABLE II. HEATSINK GEOMETRY RANGES

| | | | |
|---|---|---|---|
| **Sink length ($L_s$)** | [50-200] mm | **Fin height ($h_f$)** | [10-$f_w$] mm |
| **Sink width ($w_s$)** | Fan width ($f_w$) | **Fin width ($t_f$)** | [1-20] mm |
| **Baseplate thickness ($d_s$)** | 3 mm | **Fin number ($n_f$)** | [3-50] |

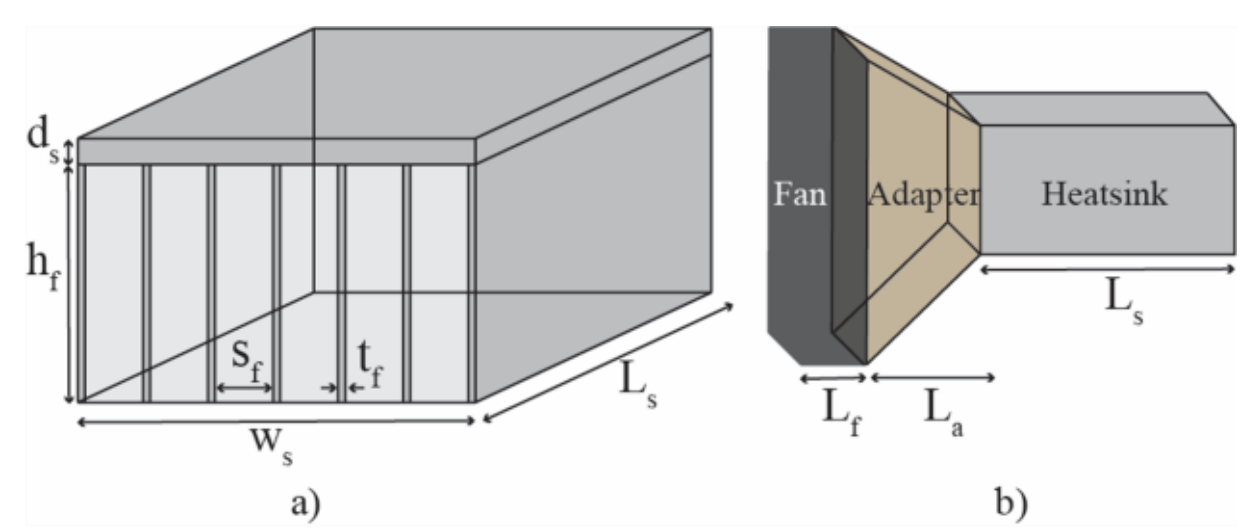

Fig. 3. (a) Detailed heatsink geometry specifying key design optimization dimensions. (b) Cooling system featuring a fan, and an adapter, and a straight-fin aluminum heat sink.

**Heatsink Cost Data:** A dataset enumerating various aluminium heatsink options sourced from Mouser as of March 1, 2024, including their costs and weights.

**Fan List:** A list of fans given in Table III, detailing their specifications to be iteratively evaluated in the optimization.

TABLE III. FAN LIST CONSIDERED FOR THE OPTIMIZATION

| Part Number | Size(HxWxL) | Air Flow | $Cost_{50}$ |
|---|---|---|---|
| 9GA0412P7G001 | 40x40x15 mm | 0.36 m3/min | $13.33 |
| 9GA0412P3J01 | 40x40x28 mm | 0.67 m3/min | $14.97 |
| 04028DA-12V-A6-KG | 40x40x28 mm | 1.13 m3/min | $20.03 |

*b) Calculation of Maximum Heatsink to Ambient Thermal Resistance ($R_{th,s-amb,max}$)*

The design constraint, $R_{th,s-a,max}$, central to the cooling system optimization, is calculated as:

$$R_{th,s\text{-}a,max} = \frac{T_{j,max} - P_{loss,sw,max} * (R_{th,j\text{-}c} + R_{th,c\text{-}s}) - T_{amb}}{P_{loss,tot,max}} \quad (5)$$

Here, $R_{th,j-c}$ is the junction to case thermal resistance given in the datasheet. The $R_{th,c-s}$ denotes the thermal resistance between semiconductor case and heatsink, determined by thermal interface material. The $P_{loss,tot,max}$ is the maximum total loss of the switches mounted on the same heatsink.

*c) Iterative Evaluation Steps*

The optimization methodology iteratively evaluates each fan from the specified list against a spectrum of heatsink geometries within defined parameter ranges. For each unique fan-heatsink pairing, the process includes:

**Volume Flow and Heatsink-to-Thermal Resistance ($R_{th,s-a}$) Calculation:** The system calculates the volume flow and the $R_{th,s-a}$ for each unique fan-heatsink pairing using analytical equations from [8].

**Configuration Validation:** For the solutions where $R_{th,s-a} < R_{th,s-a,max}$, the volume, weight, and cost of the heatsink are then computed using the following relations:

$$V_{hs} = d_s * L_s * w_s + (n_f + 1) * h_f * L_s * t_f \quad (6)$$

$$m_{hs} = V_{hs} * \rho_{Al} \quad (7)$$

$$C_{hs} = m_{hs} * \delta_{hs} \quad (8)$$

Here, $\delta_{hs}$ represents the cost per unit weight in \$/kg, determined by examining the heatsink cost data, and $\rho_{Al}$ is the density of aluminum (2.7 g/cm^3). Incorporating the fan cost from Table III, sourced from Mouser as of March 1, 2024, we record the total cost for each configuration for later evaluation.

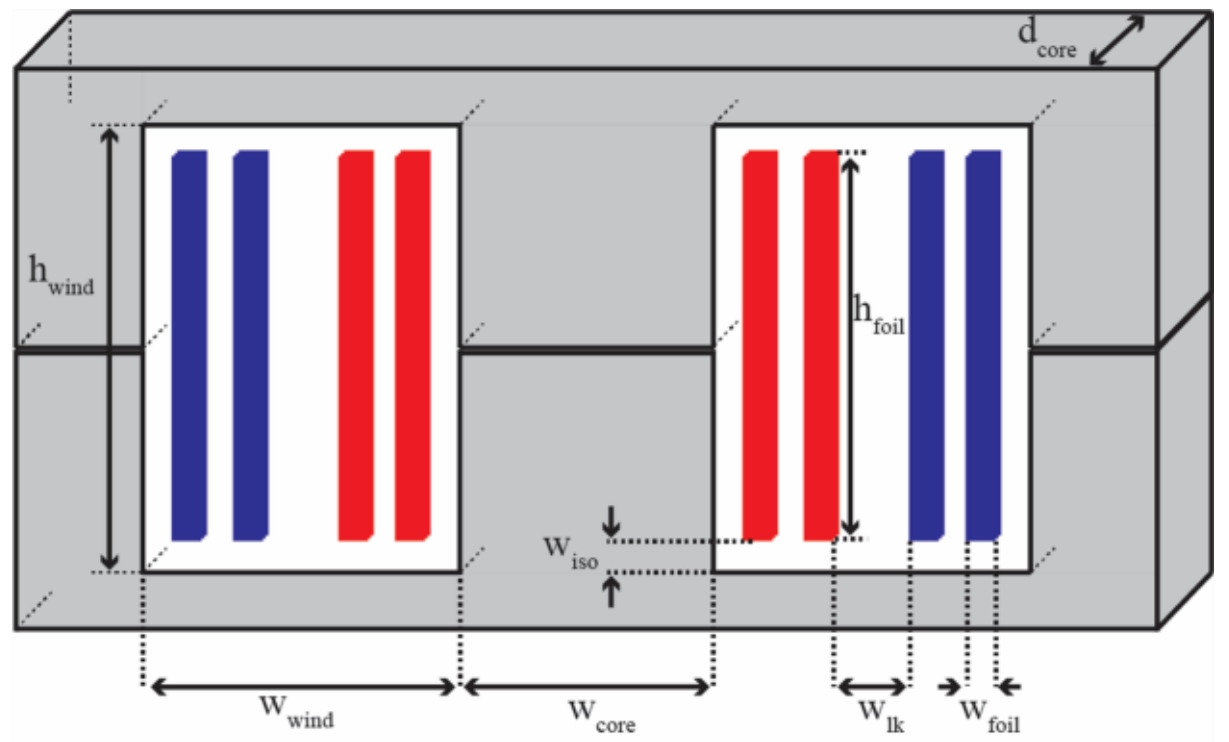

Fig. 4. (a) Detailed transformer geometry specifying key design optimization dimensions.

*d) Optimal Configuration Determination*

The process results in comparing the stored configurations to identify the most cost-efficient choice that also satisfies thermal performance criteria.

*3) Transformer* Model

This phase, as outlined in the overall optimization flow chart (Fig.1), systematically evaluates various design parameters to identify configurations that deliver the highest efficiency and the lowest cost, while adhering to all necessary operational constraints.

*a) Definition of Input Parameters*

Beside the parameters previously defined such as $T_{tx,max}$, $T_{amb}$, and the DAB converter's operational conditions $P_{conv,max}$, $I_{conv,max}$, $V_{in}$, $V_{out}$, n, $f_{sw}$, $L_{lk}$, this phase integrates following additional details:

**Core Data:** Details the core's architecture, material, and key characteristics like $B_{sat}$ and Steinmetz parameters, vital for subsequent computations. The study opts for an EE core structure with N87 ferrite material to precisely estimate leakage inductance. Cost information is derived from the latest Mouser listings as of March 1, 2024.

**Winding Data:** Includes information on the winding configuration, conductor material, and essential attributes such as density, resistivity, and temperature coefficient, necessary for ensuing calculations. Copper foil winding is selected due to their superior copper fill factor and thermal performance despite the slight increase in eddy current losses. Cost data is taken from the latest copper foil prices on [9] as of March 1, 2024.

**Design space:** Includes variables related to core geometry and winding configurations, as detailed in Table IV and showed in Fig. 4, essential for exploring different design possibilities and their impact on the transformer's performance.

TABLE IV. TRANSFORMER DESIGN PARAMATERS

|  |  |  |  |
|---|---|---|---|
| **Core width ($w_{core}$)** | [10-100] mm | **Leakage width ($w_{lk}$)** | [10-50] mm |
| **Core depth ($d_{core}$)** | [10-200] mm | **Foil width ($w_{foil}$)** | [0.1-2] mm |
| **Window height ($h_{wind}$)** | [10-200] mm | **Primary Turn (Np)** | [3-50] |

*b) Iterative Evaluation Steps*

At this phase, the goal is to thoroughly assess each design within the specified design space, using the core and winding information to systematically evaluate transformer configurations. This iterative process excludes any design that does not meet the predefined standards, advancing only feasible options for in-depth examination and optimization.

**Calculate and Evaluate $B_{max}$ and $L_{lk,tx}$:** For each potential transformer design, maximum magnetic flux density ($B_{max}$) and leakage inductance ($L_{lk,tx}$) is calculated, ensuring that $B_{max}$ stays below the core material's saturation ($B_{sat}$) and $L_{lk,tx}$ is within 10% of the specified leakage inductance, $L_{lk}$. Any design not complying with these parameters is excluded from further analysis.

**Calculate Core Loss:** The core losses ($P_{core}$) in this study are calculated by using the improved generalized Steinmetz equation as described in [10] which allows to determine the core losses also for piece-wise linear waveforms.

**Calculate Winding Loss:** Winding losses in magnetic components, especially at higher switching frequencies, are mainly dictated by skin and proximity effects that amplify ohmic losses through non-uniform current distribution. The winding power losses per unit length ($P_{wind}$) in foil conductors are calculated as described in [11].

**Calculate and Evaluate Temperature:** Along with the calculated core and winding losses, a detailed thermal model is essential to accurately compute the temperature distribution within the transformer. This research adopts the methodology from [12], providing a comprehensive structure for evaluating temperatures at critical locations. It considers heat generation from conduction, convection, and radiation, differentiating between core and winding loss impacts. By employing thermal resistance values, the model accurately forecasts steady-state temperatures, verifying they comply with the maximum temperature limit $T_{tx,max}$.

**Cost Calculation:** Once it's confirmed that the design parameters meet the predefined constraints, this phase estimates the costs for the core and windings for each feasible set of parameters.

Core Cost ($C_{core}$): Calculated from the core volume ($V_{core}$) using the formula:

$$V_{core} = 2 * w_{core} * d_{core} * (h_{wind} + w_{core} + w_{wind}) \quad (9)$$

$$C_{core} = V_{core} * \delta_c \quad (10)$$

Here, $\delta_c$ is the cost coefficient, expressed in $\$/m^3$, which is derived from a curve-fitting analysis of the costs associated with N87 E cores as obtained from the core data.

Winding Cost ($C_w$): Derived from the total volume of primary and secondary windings:

$$V_w = l_{w,P} * A_{w,P} + l_{w,S} * A_{w,S} \quad (11)$$

$$C_w = V_w * \delta_w \quad (12)$$

where $l_{w,P}$ and $l_{w,S}$ represent the mean length for primary and secondary windings, and $A_{w,P}$ and $A_{w,S}$ are their respective cross-sectional areas. The cost coefficient $\delta_w$, also expressed in $\$/m^3$, is determined using curve fitting analysis on cost of copper foils previously indicated in winding data.

*c) Store Solutions*

In the final phase, all potential solutions that have successfully passed the previous evaluations, along with their associated cost and loss data, are aggregated and forwarded to the optimization synthesis block.

*4) Capacitor Model*

Due to the wide output voltage range operation, output capacitors might be subjected to high-frequency ripple currents with significant amplitudes. These currents can significantly influence the capacitor's loss, volume, and cost, impacting the overall system design. Moreover, to maintain system stability, it is essential to limit the voltage ripple across the capacitor to a maximum allowable variation, denoted as $\Delta V_{c,max}$. This threshold is set to maximum 5V for any voltage operation from 150 V to 1000 V. The necessary output capacitance value, $C_o$, is determined as:

$$C_o = \frac{1}{4\Delta V_{c,max}} \int_0^{T_{sw}} i_c(t)dt \tag{13}$$

Here, $i_c(t)$, shown in Fig. 2.a represents the instantaneous capacitor current, calculated as the difference between the rectified secondary current $i_{s,r}(t)$ and the ouput current $i_o(t)$.

Following the capacitance requirement identification, the appropriate capacitors is selected. Film capacitors were chosen for their favourable characteristics in handling the described conditions. Cost estimation ( $C_{cap}$ ) for these capacitors is conducted using a curve fitting method based on rated capacitance ($C_c$) and voltage($V_c$), as outlined in reference [13]. The method yields a cost function expressed as:

$$C_{cap} = a_{1,C} * C_c + a_{2,C} * V_c + a_{3,C} \tag{14}$$

Furthermore, power losses ($P_{cap}$) within film capacitors, predominantly attributed to their equivalent series resistance ($R_{ESR}$) , are approximated by:

$$P_{cap} = R_{esr} * I_{c,rms}^2 \tag{15}$$

To accurately assess these losses, the $R_{ESR}$ estimation formula, tailored from [13], is:

$$R_{esr} = a_{1,R} * \frac{1}{C_c} + a_{2,R} * \frac{1}{C_c * V_C} + a_{3,R} \tag{16}$$

The empirical coefficients $a_{1,C}, a_{2,C}, a_{3,C}$ in (14) and $a_{1,R}, a_{2,R}, a_{3,R}$ in (16) are extracted from the curve fitting method by using the 1.2kV Kemet C4AQ-M capacitor series and their associated costs on Mouser as of March 1, 2024.

*G. Optimization Synthesis*

The optimization synthesis involves following three main tasks:

**Data Integration:** The outputs from the brute force algorithm, which include cost and average efficiencies for each configuration, are collected and integrated. This comprehensive dataset forms the basis for evaluating each configuration's performance.

**Normalization and Scaling:** The collected data are normalized and scaled to 350 kW for a fair comparison among different configurations. This process adjusts for differences in power ratings and component counts, making all configurations comparable on a per-unit basis. By doing so, we eliminate any bias toward configurations with inherently more or fewer modules. The power/cost ratio was then calculated by dividing the total power output by the associated cost for each configuration, providing a metric that captures the efficiency relative to the cost.

**Pareto Front Evaluation:** The Pareto front concept is used to identify the set of optimal solutions where improvements in one objective (e.g., efficiency) cannot be made without compromising another (e.g., cost).

## III. Optimization Results

In the optimization synthesis block, we collected data by exploring various combinations of design parameters and calculating the resulting efficiency and power/cost ratio. Figure 5 presents this collected data, with color maps indicating different design variables: switching frequency (fsw) in Fig 5.a, the number of modules (N) in Fig. 5.b, and the transformer turns ratio (n) in Fig. 5.c. These color maps help visualize the general trends of these variables on efficiency and the power/cost ratio.

However, when examining Fig. 5.a to Fig. 5.c, the interdependencies between efficiency, power/cost ratio, and the design variables become apparent, revealing the complexity of these trade-offs. The overlapping data points and varying color intensities make it difficult to isolate the effect of individual design variables purely from visual inspection.

To address this complexity, we applied polynomial regression methods to model and understand the relationships between the design variables (fsw, N, n) and the performance metrics (efficiency and cost). The polynomial regression plots (Fig. 5.d to Fig. 5.i) illustrate the non-linear relationships, revealing the following key insights:

**Switching Frequency (fsw):** Higher switching frequencies are generally associated with lower efficiencies (Fig. 5.d) and a slight increase in the power-cost ratio (Fig. 5.g).

**Number of Converter Modules (N):** There is little or no clear relationship between the number of modules and efficiency (Fig. 5.e). However, the power-cost ratio decreases as the number of modules increases (Fig. 5.h).

**Transformer Turns Ratio (n):** There is an optimal range for the turns ratio where efficiency is maximized (Fig. 5.f) and the power-cost ratio is balanced (Fig. 5.i). This suggests that neither increasing nor decreasing the turns ratio straightforwardly improves performance; instead, a balanced approach within an optimal range is necessary.

Since there is a tradeoff between efficiency and power/cost ratio, choosing the optimal design depends on the designer priorities. Some may prioritize the highest efficiency, while others may prioritize the lowest cost. The concept of the Pareto front is useful here, as it represents the set of optimal solutions where any improvement in one objective would lead to a deterioration in another.

In this study, we applied an additional constraint, selecting the lowest cost designs that also provide at least 95% efficiency across the specified output voltage range at rated power. From this filtered dataset, the selected design, marked with a star, employing seven 50 kW DAB modules with the switching frequency of 30 kHz and the transformer turns ratio of 0.9 stands out for meeting this efficiency threshold while also optimizing cost.

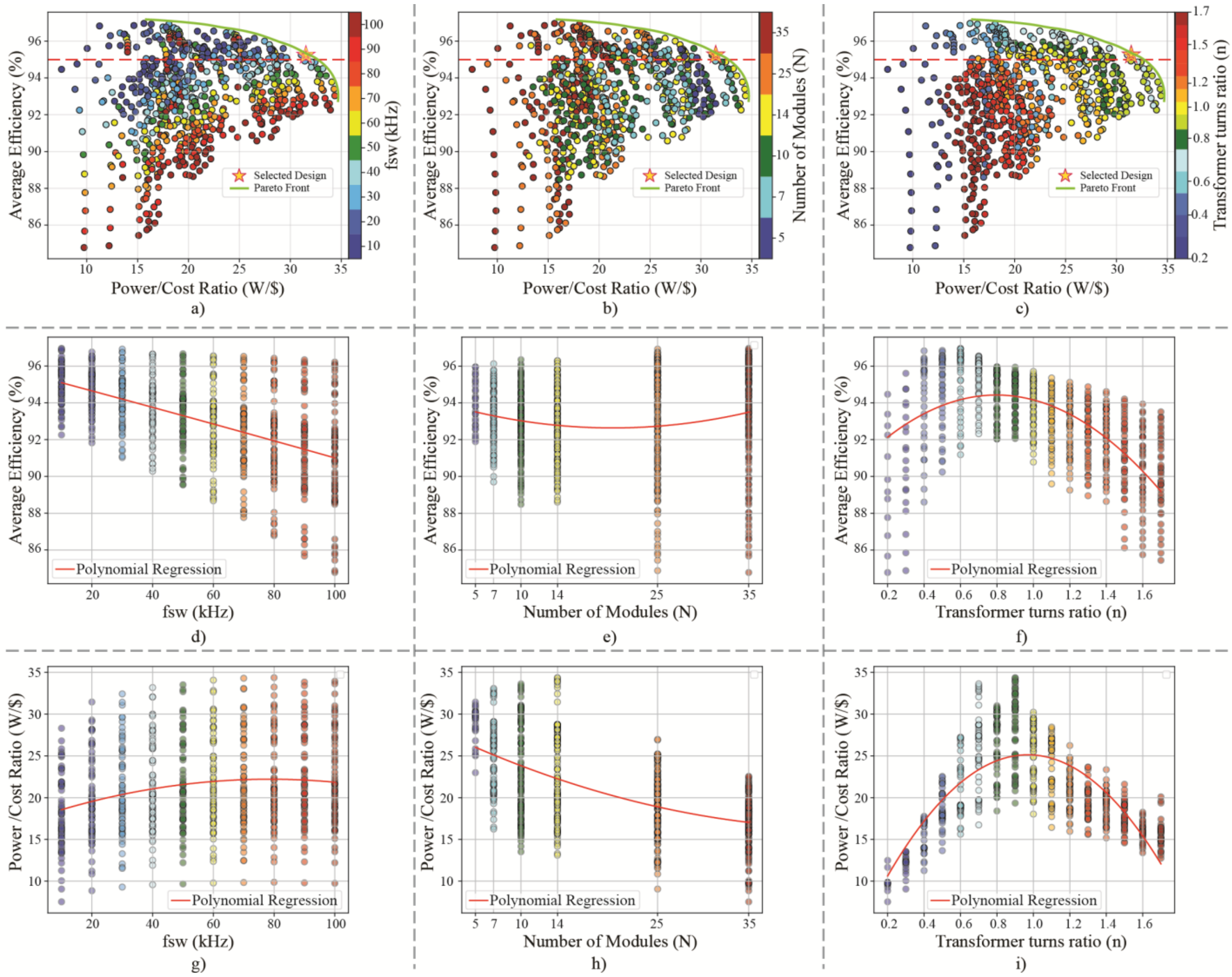

Fig.5. Analysis of design parameters impact on DAB converter performance. (a-c) Average efficiency vs. power/cost ratio with color maps indicating switching frequency (fsw), number of modules (N), and transformer turns ratio (n), respectively. The Pareto front represents optimal trade-offs. (d-f) Polynomial regression of efficiency vs. fsw, N, and n, respectively. (g-i) Polynomial regression of power-cost ratio vs. fsw, N, and n, respectively.

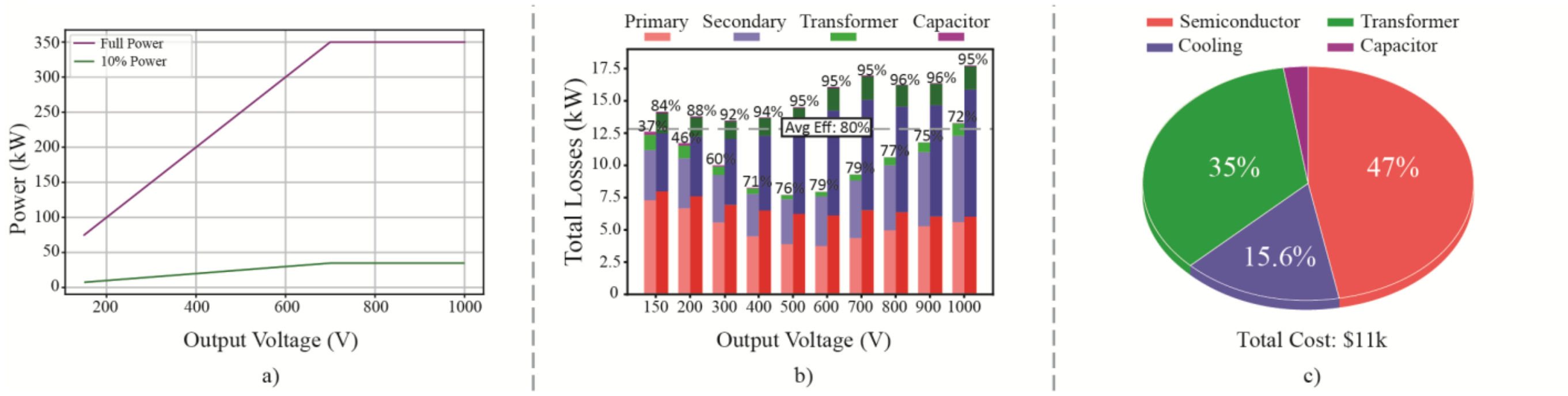

Fig.6. (a) Maximum and 10% output power across the specified output voltage range (150 V to 1000 V). (b) Efficiency comparison at light load (10% of full power, depicted with lower color intensity) and maximum power across the output voltage range. (c) Total cost breakdown of the main components for the full 350 kW charger.

The performance of this design, as showcased in Fig. 6, highlights several key aspects. Fig. 6.a shows the maximum and 10% output power levels that the DC fast charger needs to supply across the specified voltage range (150 V to 1000 V). These power levels are used for efficiency calculations in Fig. 6.b, which demonstrates that the design maintains an efficiency above 95% at full power across varying operational conditions. However, the average efficiency across all conditions, including both 100% and 10% of full power within a wide voltage range, falls to around 80%. This reduction is primarily due to the Single-Phase Shift modulation technique used, which causes efficiency losses under wide voltage and load variations. Fig. 6.c provides a detailed breakdown of the main contributors to the converter's cost, with the largest part coming from the semiconductor, followed by costs for transformer, heatsinks, and capacitors.

## IV. Conclusion

This study presents the comprehensive optimization framework for the design parameters of a Dual Active Bridge converter tailored for 350 kW DC fast chargers, with a focus on balancing efficiency and cost. The findings indicate that the optimal configuration consists of seven 50 kW DAB modules operating at a switching frequency of 30 kHz with a transformer turns ratio of 0.9. This configuration achieves an efficiency of over 95% at rated power across the wide output voltage range. A key limitation identified is the efficiency drop at light loads, attributed to the SPS modulation strategy used. Future work should investigate alternative modulation techniques and conduct experimental validation to further optimize DAB converter performance across all load conditions.

## V. Acknowledgements

This work has received funding from the Europe-an Union's Horizon Europe research and innovation programme under the Marie Sklodowska-Curie Doctoral Networks grant agreement No 101072414 (E2GO).

## References

[1] "Global EV Outlook 2023," International Energy Agency (IEA), Paris, France, Tech. Rep.,2020. [Online]. Available: https://www.iea.org /reports/global-ev-outlook-2023. [Accessed: Mar. 14, 2024].

[2] C. Jung, "Power Up with 800-V Systems: The benefits of upgrading voltage power for battery-electric passenger vehicles," in IEEE Electrification Magazine, vol. 5, no. 1, pp. 53-58, March 2017.

[3] S. Rivera et al., "Electric Vehicle Charging Infrastructure: From Grid to Battery," in IEEE Industrial Electronics Magazine, vol. 15, no. 2, pp. 37-51, June 2021.

[4] D. Lyu, C. Straathof, T. B. Soeiro, Z. Qin and P. Bauer, "ZVS-Optimized Constant and Variable Switching Frequency Modulation Schemes for Dual Active Bridge Converters," in IEEE Open Journal of Power Electronics, vol. 4, pp. 801-816, 2023.

[5] O. Zayed, A. Elezab, A. Abuelnaga and M. Narimani, "A Dual-Active Bridge Converter With a Wide Output Voltage Range (200–1000 V) for Ultrafast DC-Connected EV Charging Stations," in IEEE Transactions on Transportation Electrification, vol. 9, no. 3, pp. 3731-3741, Sept. 2023.

[6] C. Wei, Z. Hu, F. Zhang, J. Shao and A. Narain, "A SiC Based 30kW Three-Phase Interleaved LLC Resonant Converter for EV Fast Charger," PCIM Europe 2022; International Exhibition and Conference for Power Electronics, Intelligent Motion, Renewable Energy and Energy Management, Nuremberg, Germany, 2022, pp. 1-6.

[7] Charging Interface Initiative e.V., "DC CCS Power Classes V7.2," Dec. 9, 2021. [Online].Available:https://www.charin.global/media/ pages/technology /knowledge-base/c6574dae0e-1639130326/charin_ dc_ccs_power_classes.pdf. [Accessed: Mar. 14, 2024].

[8] C. Gammeter, F. Krismer and J. W. Kolar, "Weight optimization of a cooling system composed of fan and extruded fin heat sink," 2013 IEEE Energy Conversion Congress and Exposition, Denver, CO, USA, 2013, pp. 2193-2200.

[9] Basic Copper, "Copper Sheet & Rolls," Basic Copper. [Online]. Available: https://basiccopper.com/copper-sheet--rolls.html. [Accessed: Mar. 14, 2024].

[10] K. Venkatachalam, C. R. Sullivan, T. Abdallah and H. Tacca, "Accurate prediction of ferrite core loss with nonsinusoidal waveforms using only Steinmetz parameters," 2002 IEEE Workshop on Computers in Power Electronics, 2002. Proceedings., Mayaguez, PR, USA, 2002, pp. 36-41

[11] J. Mühlethaler, "Modeling and multi-objective optimization of inductive power components", Ph.D. dissertation, ETH Zürich,2012.

[12] M. Bahmani, "Design and Optimization Considerations of Medium-Frequency Power Transformers in High-Power DC-DC Applications," PhD thesis, Chalmers University of Technology, Gothenburg, 2016.

[13] D. Christen, "Analysis and Performance Evaluation of Converter Systems for EV-Ultra-Fast Charging Stations with Integrated Grid Storage" , Ph.D. dissertation, ETH Zürich,2017.

[14] A. Blinov, D. Vinnikov, and T. Jalakas, "Loss Calculation Methods of Half-Bridge Square-Wave Inverters", ELEKTRON ELEKTROTECH, vol. 113, no. 7, pp. 9-14, Sep. 2011.